\documentclass[a4paper,twocolumn,11pt, longbibliography,accepted=2022-01-11]{quantumarticle}

\pdfoutput=1
\usepackage[utf8]{inputenc}
\usepackage[english]{babel}
\usepackage[T1]{fontenc}
\usepackage{amsmath}
\usepackage{hyperref}
\usepackage{cleveref}
\usepackage{svg}
\usepackage[numbers,sort&compress]{natbib}

\usepackage{tikz}
\usepackage{lipsum}
\usepackage{natbib}
\usepackage{graphicx, adjustbox}
\usepackage{pythonhighlight}
\usepackage{subcaption}
\usepackage{physics}
\usepackage{qcircuit}
\usepackage{tcolorbox} 

\usepackage{listings}
\usepackage{xcolor}

\lstnewenvironment{numbered_python}[1][]{\lstset{style=mypython,numbers=left}}{}

\definecolor{codegreen}{rgb}{0,0.6,0}
\definecolor{codegray}{rgb}{0.5,0.5,0.5}
\definecolor{codepurple}{rgb}{0.58,0,0.82}
\definecolor{backcolour}{rgb}{0.95,0.95,0.92}
\lstdefinestyle{mystyle}{
    backgroundcolor=\color{backcolour},   
    commentstyle=\color{codegreen},
    keywordstyle=\color{magenta},
    numberstyle=\tiny\color{codegray},
    stringstyle=\color{codepurple},
    basicstyle=\ttfamily\footnotesize,
    breakatwhitespace=false,         
    breaklines=true,                 
    captionpos=b,                    
    keepspaces=true,                 
    numbers=none,                    
    numbersep=5pt,                  
    showspaces=false,                
    showstringspaces=false,
    showtabs=false,                  
    tabsize=2
}
\newcommand{\beq}{\begin{eqnarray}}
\newcommand{\eeq}{\end{eqnarray}}
\renewcommand{\eqref}[1]{\mbox{Eq.~(\ref{#1})}}
\renewcommand{\hat}[1]{#1}

\crefname{figure}{Figure}{Figures}
\crefname{equation}{Eq.}{Equations}
\crefname{section}{Section}{Sections}

\begin{document}

\title{Pulse-level noisy quantum circuits with QuTiP}
\author{Boxi Li}
\affiliation{Peter Gr\"unberg Institute - Quantum Control (PGI-8), Forschungszentrum J\"ulich GmbH, D-52425 J\"ulich, Germany}
\orcid{0000-0002-2733-7186}
\email{b.li@fz-juelich.de}
\author{Shahnawaz Ahmed}
\email{shahnawaz.ahmed95@gmail.com}
\orcid{0000-0003-1145-7279}
\affiliation{Department of Microtechnology and Nanoscience, Chalmers University of Technology, 412 96 Gothenburg, Sweden}
\author{Sidhant Saraogi}
\affiliation{Department of Computer Science, Georgetown University, 3700 O St NW, Washington, DC 20057, United States
}
\author{Neill Lambert}
\affiliation{Theoretical Quantum Physics Laboratory, RIKEN Cluster for Pioneering Research, Wako-shi, Saitama 351-0198, Japan}
\orcid{0000-0001-7873-0773}
\author{Franco Nori}
\orcid{0000-0003-3682-7432}
\affiliation{Theoretical Quantum Physics Laboratory, RIKEN Cluster for Pioneering Research, Wako-shi, Saitama 351-0198, Japan}
\affiliation{RIKEN Center for Quantum Computing (RQC), 2-1 Hirosawa, Wako-shi, Saitama 351-0198, Japan}
\affiliation{Department of Physics, University of Michigan, Ann Arbor, Michigan 48109-1040, USA}
\author{Alexander Pitchford}
\affiliation{Department of Mathematics, Aberystwyth University, Penglais Campus, Aberystwyth, SY23 3BZ, Wales, United Kingdom}
\orcid{0000-0002-4717-2921}
\author{Nathan Shammah}
\affiliation{Unitary Fund, Walnut, California 91789, USA}
\email{nathan@unitary.fund}
\orcid{0000-0002-8775-3667}


\maketitle
\begin{abstract}
The study of the impact of noise on quantum circuits is especially relevant to guide the progress of Noisy Intermediate-Scale Quantum (NISQ) computing.
In this paper, we address the pulse-level simulation of noisy quantum circuits with the Quantum Toolbox in Python (QuTiP). 
We introduce new tools in \texttt{qutip-qip}, QuTiP's quantum information processing package. These tools simulate quantum circuits at the pulse level, leveraging QuTiP's quantum dynamics solvers and control optimization features. We show how
quantum circuits can be compiled on simulated processors, with control pulses acting on a target Hamiltonian that describes the unitary evolution of the physical qubits. Various types of noise can be introduced based on the physical model, e.g., by simulating the Lindblad density-matrix dynamics or Monte Carlo quantum trajectories. In particular, the user can define environment-induced decoherence at the processor level and include noise simulation at the level of control pulses. We illustrate how the Deutsch-Jozsa algorithm is compiled and executed on a superconducting-qubit-based processor, on a spin-chain-based processor and using control optimization algorithms. We also show how to easily reproduce experimental results on cross-talk noise in an ion-based processor, and how a Ramsey experiment can be modeled with Lindblad dynamics. Finally, we illustrate how to integrate these features with other software frameworks.

\end{abstract}

\section{Introduction}

Quantum computation and quantum algorithms are deemed to be able to complete tasks that would be harder or impossible to achieve with classical resources. However, noise on quantum hardware significantly influences its performance, limiting large-scale applications. Currently, we are in the so-called noisy intermediate-scale quantum (NISQ) computing era \cite{Preskill2018}. Before we reach the regime of quantum error correction (QEC)~\cite{nielsen2002quantum}, quantum algorithms will suffer from quantum and classical noise, e.g., decoherence and noise in classical control signals. Both types of noise lead to errors in the computation and therefore determine the performance of a quantum algorithm. Hence, a realistic simulation of a quantum algorithm needs to incorporate these different types of noise, which can depend strongly on the type of qubit technology \cite{buluta2011natural}.

A modern quantum algorithm typically includes both classical and quantum parts \cite{bharti2021noisy}. The former can include classical variational subroutines, while the latter is usually represented by a quantum circuit, consisting of a number of gates applied on a quantum state. Many software projects provide the simulation of such circuits including PyQuil \cite{smith2016practical,Karalekas_2020}, Qiskit \cite{Qiskit}, Cirq \cite{cirq}, ProjectQ \cite{Steiger2018}, and PennyLane \cite{Bergholm2018}, among others~\cite{Fingerhuth_2018,heim2020quantum}.
However, within these approaches, noise is usually modelled as an additional layer on top of ideal quantum gates, e.g., probabilistically inserting random Pauli gates or a list of Kraus operators to describe a noisy quantum channel.

To improve the performance of a quantum circuit on noisy hardware, it is useful to also perform optimization at the level of control pulses based on the quantum dynamics of the underlying hardware.
For this purpose, open-source software packages have been developed to map quantum circuits to control pulses on hardware, allowing for fine-tuning and calibration of the control pulses, such as \texttt{qiskit.pulse}~\cite{Alexander2020}, \texttt{qctrl-open-controls}~\cite{Ball_2020} and \texttt{Pulser}~\cite{pasqal2020}.
Recently, Qiskit also launched the project \texttt{qiskit-dynamics} to support solving time-dependent quantum systems, connected with \texttt{qiskit.pulse}. The project is still in the early stages of development.

In the realm of simulation,
one of the earliest, and most widely used Python packages to simulate quantum dynamics is the Quantum Toolbox in Python, QuTiP \cite{Johansson12,Johansson13}. QuTiP provides useful tools for handling quantum operators and simplifies the simulation of a quantum system under a noisy environment by providing a number of solvers, such as the Lindblad master equation solver. An ecosystem of software tools for quantum technology is growing around it~\cite{Shammah_2018,Goerz_2019_SciPost,lambert2019modelling,lambert2020bofinheom,qopt,Khaneja2005,pasqal2020,horn2021,Alexander2020,Scqubits}. Hence, it is a natural base to start connecting the simulation of quantum circuits and the time evolution of the quantum system representing the circuit registers. 
At the cost of more computing resources, simulation at the level of time evolution allows noise based on the physical model to be included in the realistic study of quantum circuits. 

\paragraph{Summary of results}
In this paper, we illustrate how the new tools in \texttt{qutip-qip}\footnote{\url{https://github.com/qutip/qutip-qip}} can be used to bridge the gap between the gate-level circuit simulation and the simulation of quantum dynamics following the master equation for various hardware models. While a quantum circuit representation and a few specific Hamiltonian models have been available in QuTiP for some time, in this paper, we bridge them with QuTiP solvers and build a pulse-level simulation framework, allowing the simulation of noisy circuits.

\begin{figure}[th!]
  \centering
  \includegraphics[width=\linewidth]{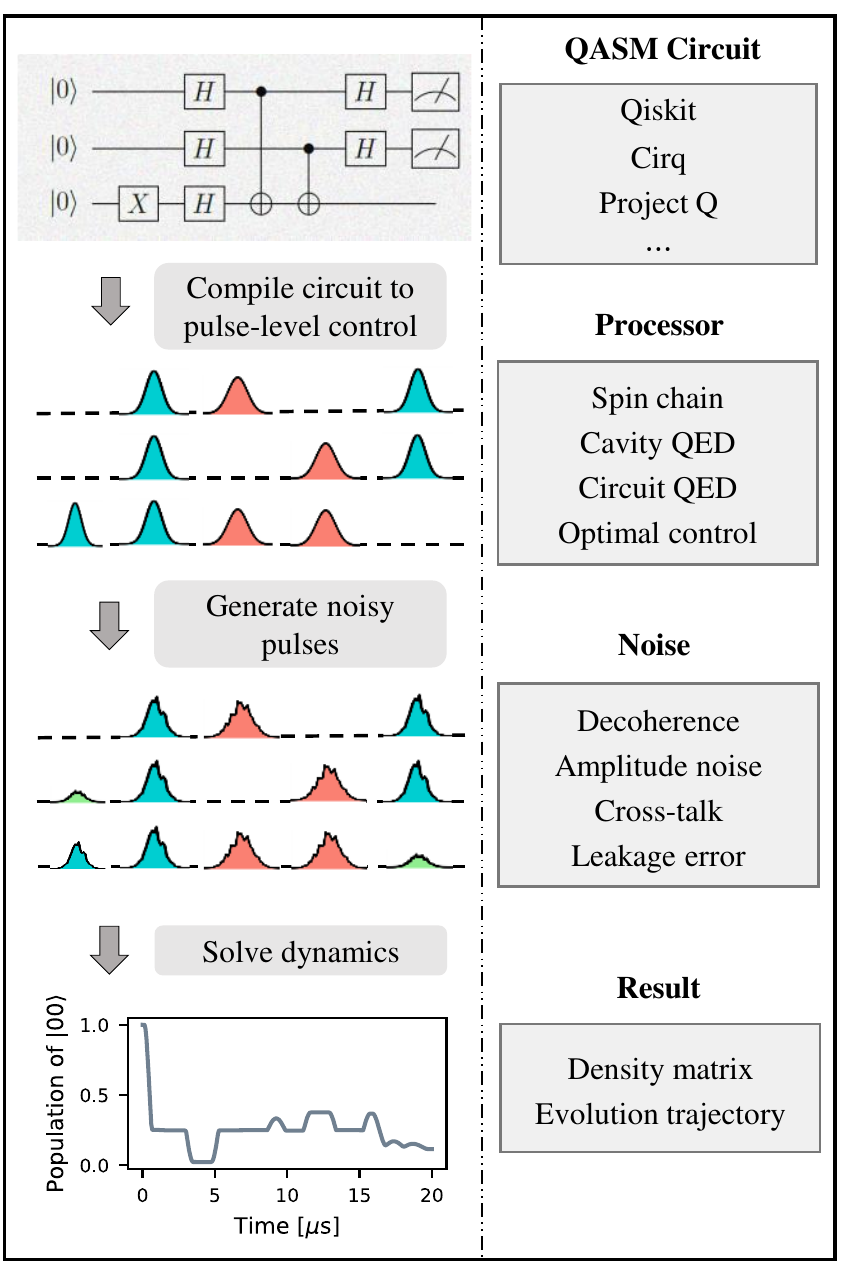}
  \caption{
    Illustration of the workflow of the pulse-level noisy quantum circuit simulation.
    It starts from a quantum circuit defined in QuTiP or imported from other libraries through the QASM format.
    Based on the hardware of interest, the circuit is then compiled to control pulse signals for each control Hamiltonian (blue for single-qubit gates and red for two-qubit gates in the figure).
    Next, a representation of the time evolution, including various types of noise, is generated under the description of the master equation.
    In the last step, the QuTiP solver is employed to solve the dynamics.
    The solver returns the final result as well as the intermediate state information on demand.
    Both the final and the intermediate quantum states can be recorded, as illustrated by the plot showing the population of the $\ket{00}$ state, with the third qubit traced out.
    This plot is the same as \cref{fig:population} and will be explained later in detail.
    The control signals in the figure are for illustration purposes only while the real compiled pulses on a few predefined hardware models are shown in \cref{fig:compiled_pulse}.
  }
  \label{fig:illustration}
\end{figure}

Provided a Hamiltonian model and a map between the quantum gate and control pulses, we show how these new tools in \texttt{qutip-qip} can be used to compile the circuit into the native gates of a given hardware, how to generate the physical model described by control pulses and how to use QuTiP's dynamical solvers to obtain the full-state time evolution, as shown in \cref{fig:illustration}.

A number of example hardware models are available in the software package -- a spin qubit processor, a cavity-QED device, a superconducting qubit model -- while in general the users are provided with the freedom to define their own devices of choice. 

In addition to a predefined map between gates and pulses for each model, optimal control algorithms in QuTiP can also be used to generate control pulses. Moreover, we demonstrate how various types of noise, including decoherence induced by the quantum environment and classical control noise, can be introduced at different layers of the simulation. Thanks to a modular code design, one can quickly extend the toolkit with customized hardware and noise models.

\paragraph{Article structure} The article is organized as follows: In Section~\ref{sec:software}, information about the software installation and specifics is given. In Section \ref{sec:qc-openqs}, we briefly present the background concepts of quantum circuits at the gate level, the continuous-time pulse-level description for circuits, open quantum systems theory and the tools present in \texttt{qutip} and \texttt{qutip-qip} to represent and simulate open quantum systems. Section \ref{sec:framework} contains the main novel results and new software features: therein, we illustrate in detail the novel architecture of the pulse-level quantum-circuit simulation framework in \texttt{qutip-qip} and the available modelling of quantum devices and noise. In Section \ref{sec:qasm}, we show how these features can be integrated with other software through importing external quantum circuits using the QASM format. We conclude in Section \ref{sec:conclusion}. 

The Appendices include self-contained code examples: \Cref{sec:dj} contains the full code for the Deutsch-Jozsa algorithm simulation; \Cref{sec:qft} presents the simulation of a 10-qubit quantum Fourier transform (QFT) algorithm using the spin chain model; \Cref{sec:customization appendix} shows how to customize the physical model of a processor with noise.
More examples can also be found in QuTiP tutorials\footnote{\url{http://qutip.org/tutorials.html} under the section Quantum Information Processing}.

\section{Software information}
\label{sec:software}
The tools described here are part of the QuTiP project \cite{Johansson12, Johansson13}.
The \texttt{qutip-qip} package builds upon what was once a module of QuTiP, \texttt{qutip.qip}.
Usage and installation has not significantly changed for the end user, who can easily install the package from the Python package index (PyPI) distribution with
\\
\begin{lstlisting}[language=Python, caption=Installing qutip-qip]
pip install qutip-qip
\end{lstlisting}

The \texttt{qutip-qip} package has the core \texttt{qutip} package as its main dependency. This means that it also builds upon the wider Python scientific open source software stack, including NumPy \cite{harris2020array} and SciPy \cite{virtanen2020scipy}, and optionally Matplotlib \cite{hunter2007matplotlib} and Cython \cite{behnel2010cython}. 
\texttt{qutip-qip} is a software developed by many contributors \cite{qutipqip}.

The \texttt{qutip-qip} package is developed with the best practices of open-source software development and scientific software. The codebase is hosted on Github and new code contributions are reviewed by the project maintainers. The license is the BSD three-clause license (also known as BSD 2.0 or New BSD). The code is thoroughly unit tested, with tests for most objects also run on the cloud in continuous integration, on multiple operating systems. The documentation, whose code snippets and API documentation are also unit tested, is hosted online on Read The Docs (\url{https://qutip-qip.readthedocs.io/}); the documentation can also be generated locally by contributors with Sphinx by forking the \texttt{QuTiP/qutip-qip} Github repository. 

\section{Quantum circuits and open quantum dynamics}
\label{sec:qc-openqs}
In this section, we briefly review the theory of quantum circuits and their modelling on actual devices that are subject to noise. We introduce the formalism for the gate-level representation of quantum circuits, then describe the Hamiltonian description at the pulse level, and finally the open-quantum dynamics of a realistic system.  

\subsection{Quantum circuits and gate-level simulation}
\label{sec:circuit}
A quantum circuit is a model for quantum computation, where the quantum dynamics is abstracted and broken down into unitary matrices (quantum gates), which can be applied to all or only a few circuit registers.
Inherited from classical computing, the circuit registers are most often two-level systems, referred to as qubits.
The execution of a circuit on a quantum state is then given by \begin{equation}
    \ket{\psi_f} = U_K U_{K-1} \cdots U_2 U_1 \ket{\psi_i},
    \label{eq:quantum circuit}
 \end{equation}
where $\ket{\psi_i}$ and $\ket{\psi_f}$ are the initial and final state and $U_k$ with $k \in \{1,2,\cdots,K\}$ the quantum gates.

Often, the simulation of quantum circuits is implemented by representing the unitaries and quantum state as complex matrices and vectors.
The execution of a circuit is then described as matrix-vector multiplication.
We refer to this as gate-level quantum circuit simulation.
The gate-level quantum circuit description is a representation of a quantum algorithm at an abstract level before considering any physical realization to implement the algorithm \cite{nielsen2002quantum}. More general representations of hybrid quantum algorithms include the integration of classical and quantum subroutines, as for variational quantum algorithms \cite{bharti2021noisy}, their compilation and execution \cite{cross2017open,Bergholm2018,Karalekas_2020,cross2021openqasm,qcor}.  
In \texttt{qutip-qip}, this gate-level simulation can be performed with the \texttt{QubitCircuit} class, which is the Python object used to represent a quantum circuit.

In order to introduce the effects of noise, quantum states can be most generally represented by a density matrix and the idea of a quantum channel is introduced, where noise can be characterized by a set of non-unitary Kraus operators acting on the quantum states.
Many well-known channel representations of noise have been implemented in circuit simulation, such as depolarising, dephasing, amplitude damping and erasure channel.
Although the channel description is very general, noisy gate simulation based on it has two shortcomings.

First, in most implementations, noise is applied after the ideal gate unitaries, while in reality they are not separated. 
Second, although quantum channels describe the most general evolution that a quantum system can undergo, finding the representation of realistic noise in this channel form is not a trivial task.
Usually, a noise channel implemented in simulators only describes single-qubit decoherence and cannot accurately capture the complicated noisy evolution that the system undergoes.
Hence, to study the execution of circuits on noisy hardware in more detail, one needs to turn to the quantum dynamics of the hardware platform.

\label{sec:qir}

\subsection{Continuous time evolution and pulse-level description}
\label{sec:solvers}
Down to the physical level, quantum hardware, on which a circuit is executed, is described by quantum theory.
The dynamics of the system that realizes a unitary gate in \cref{eq:quantum circuit} is characterized by the time evolution of the quantum system.
For isolated or open quantum systems, we consider both unitary time evolution and open quantum dynamics.
The latter can be simulated either by solving the master equation or sampling Monte Carlo trajectories.
Here, we briefly describe those methods as well as the corresponding solvers available in QuTiP.
\subsubsection{Unitary time evolution}
\label{sec:unitary evolution}
For a closed quantum system, the dynamics is determined by the Hamiltonian and the initial state.
From the perspective of controlling a quantum system, the Hamiltonian is divided into the non-controllable drift $H_{\textnormal{d}}$ (which may be time dependent) and controllable terms combined as $H_{\textnormal{c}}$ to give the full system Hamiltonian
\begin{equation}
\label{eq:Hamiltonian}
    H(t) = H_{\textnormal{d}}(t) + H_{\textnormal{c}}(t) = H_{\textnormal{d}}(t) + \sum_j c_j(t) H_j,
\end{equation}
where the $H_j$ describe the effects of available physical controls on the system that can be modulated by the time-dependent control coefficients $c_j(t)$, by which one drives the system to realize the desired unitary gates.

The unitary $U$ that is applied to the quantum system driven by the Hamiltonian $H(t)$ is a solution to the Schrödinger operator equation
\begin{equation}
    i \hbar \frac{\partial U(t)}{\partial t}
    = H(t) U(t)
    .
\end{equation}
By choosing $H(t)$ that implements the desired unitaries (the quantum circuit) we obtain a pulse-level description of the circuit in the form of \cref{eq:Hamiltonian}. 
The choice of the solver depends on the parametrization of the control coefficients $c_j(t)$. 
The parameters of $c_j(t)$ may be determined through theoretical models or automated through control optimisation, as described later in \cref{sec:framework}.

\subsubsection{Open quantum system dynamics}
In reality, a quantum system is never perfectly isolated; hence, a unitary evolution is often only an approximation.
To consider possible interaction with the environment, one can introduce a larger Hilbert space, or reduce the overhead by effectively limiting the description to the system Hilbert space and using super-operators inducing a non-unitary dynamics (i.e., on an open system).
One way to describe the evolution of an open quantum system is by the Lindblad master equation.
It can be solved either by solving a differential equation (\texttt{qutip.mesolve}) or by Monte Carlo sampling of quantum trajectories (\texttt{qutip.mcsolve}).
Both can be chosen as a simulation back-end for the pulse-level circuit simulator.

These solvers provide an efficient simulation of open system quantum dynamics. They can describe noise models derived under the the Born-Markov Secular (BMS) approximations \cite{breuer2002theory,Lidar2019}, and more general Lindbladians, including those with time-dependent rates. For most hardware implementations these noise models are powerful and flexible enough to capture the most salient environmental noise effects.

\paragraph{Density-matrix master equation solver.}
\label{sec:lindblad}
The function \texttt{qutip.mesolve} can solve general open dynamics that can be cast in the form

\begin{align}
    \frac{\partial \hat{\rho}(t)}{\partial t} =& \mathcal{L} \hat{\rho}(t),
\label{Eq:general_mesolve}
\end{align}
where the dynamics of the ``system''
density matrix $\hat{\rho}(t)$ evolves under the action of a superoperator $\mathcal{L}$. The user can decide to provide directly the full superoperator $\mathcal{L}$, or divide the dynamics in the Hamiltonian part [\eqref{eq:Hamiltonian}] and noise terms provided by a set of collapse operators (\texttt{c\_ops}) with related rates, and \texttt{qutip.mesolve} will effectively solve Eq.~(\ref{Eq:general_mesolve}) behind the scenes. The structure of Eq.~(\ref{Eq:general_mesolve}) can be quite generic, including the possibility for time-dependent rates and collapse operators, beyond the Born–Markov and secular (BMS) approximation, however, one of the most straightforward approaches is to simulate a Lindblad master equation. A common example for a quantum circuit consisting of $N$ qubits experiencing relaxation and dephasing would be the following Lindblad master equation,
\begin{align}
    \frac{\partial \hat{\rho}(t)}{\partial t} =& 
    - i \left[\hat{H}(t), \hat{\rho}(t)\right] +   \sum_{j=0}^{N-1}\gamma_j \mathcal{D}[\hat{\sigma}^-_j] \hat{\rho}(t) \nonumber\\
    +&   \sum_{j=0}^{N-1}\frac{\gamma^D_j}{2} \mathcal{D}[\hat{\sigma}^z_j] \hat{\rho}(t),
\label{Eq:Lindblad}
\end{align}
where $\hat{H}$ is the system Hamiltonian, $\gamma_j$ is the relaxation rate of qubit $j$, $\gamma_j^D$ the pure dephasing rate of qubit $j$, $\mathcal{D}[\Gamma_n]X=\Gamma_n X \Gamma_n^\dagger-\frac{1}{2} \Gamma_n^\dagger \Gamma_n X-\frac{1}{2}X \Gamma_n^\dagger \Gamma_n$ is the Lindblad dissipator for a generic jump operator $\Gamma_n$ acting on a density matrix $X$, and $\hat{\sigma}^\alpha_j$ are Pauli operators, with $\alpha=x,y,z,+,-$.

This approach allows us to model the coexistence of pulse-level control, in the coherent Hamiltonian part, and the influence of noise. However, the density matrix description of the system introduces a quadratic overhead in memory size. If this becomes a limiting factor for a given simulation, progress can be made by employing the Monte-Carlo quantum trajectory solver, \texttt{qutip.mcsolve}.

\paragraph{Monte-Carlo quantum trajectories.}
A popular method that is alternative to the full master equation simulation is the Monte Carlo sampling with quantum trajectories. Noise is included in an effective non-Hermitian Hamiltonian, and a stochastic term is added by pseudo-random sampling. An effective Hamiltonian is continuously applied to the system, integrating the part of \cref{Eq:Lindblad} with Lindblad dissipators,   
\begin{eqnarray}
\label{Eq:MonteCarlo}
    \hat{H}_\text{eff}&=&\hat{H}(t)-\frac{i}{2}\sum_n\hat{\Gamma}_n^\dagger\hat{\Gamma}_n,
\end{eqnarray}
while the second part is determined stochastically, checking if a random number is greater than the norm of the unnormalized wavefunction. If that is the case, the quantum jump is applied, ensuring the renormalization of the wavefunction, 
\begin{eqnarray}
\label{Eq:MonteCarloFlip}
    |\psi(t+\delta t)\rangle &=&\frac{\hat{\Gamma}_n|\psi(t+\delta t)\rangle}{\sqrt{\langle \psi(t)|\hat{\Gamma}_n^\dagger \hat{\Gamma}_n|\psi(t)\rangle }}
    .
\end{eqnarray}
The advantage of the quantum trajectory approach over the density-matrix master equation solution is that one needs to handle a computational space of dimension $N$ equal to the Hilbert space, instead of its square. Additionally, the quantum-trajectory approach allows simulating the dynamics of single executions instead of the averaged dynamics from a density-matrix simulation using the master equation, which can provide further insight in processes that may be washed out when looking only at the statistical averages \cite{carmichael2009statistical,Minganti16}. A trade-off is present in the number of trajectories that need to be run to evaluate a mean path with a small standard deviation. However, the trajectories can be computed in parallel. QuTiP uses Python's multiprocessing module to benefit from multi-core computing platforms.

\label{sec:montecarlo}
\label{sec:mcsolve}
\paragraph{Other dynamical solvers.}
\label{sec:othersolvers}
QuTiP also provides solvers for other noise models and dynamics, such as the (secular and non-secular) Bloch-Redfield equation \cite{breuer2002theory}, the (non-Markovian) hierarchical equation of motion (HEOM) \cite{Tanimura_1989,lambert2020bofinheom}, and stochastic master equations. These are not currently supported for the pulse-level circuit simulation of \texttt{qutip-qip}.

\section{Pulse-level quantum-circuit simulation framework}
\label{sec:framework}

\begin{figure}
    \centering
    \includegraphics{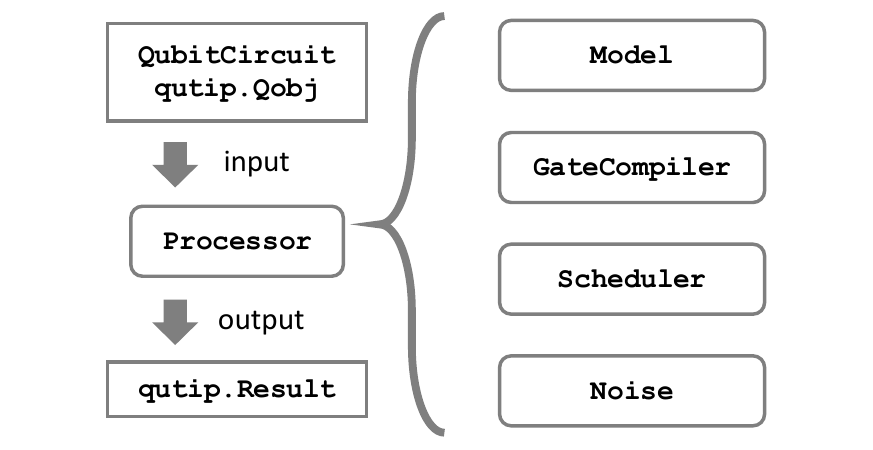}
    \caption{The structure of the simulation framework. The main interface is implemented in the class \texttt{Processor}.
    An instance of \texttt{Processor} emulates a quantum processor that takes a circuit and an initial quantum state as input and outputs the result as a \texttt{qutip.Result} object.
    From the result, one can inspect the final state of the physical qubits, as well as intermediate results during the time evolution. The \texttt{Processor} has a modular design that allows for arbitrary specifications of the underlying hardware model, compilation, scheduling gates and noise models.
    }
    \label{fig:structure}
\end{figure}

In this section, we describe the architecture of the simulation framework.
The framework aims at simplifying the simulation of noisy quantum circuits through the explicit time evolution of physical qubits using QuTiP solvers.
As illustrated in \cref{fig:structure}, the simulation is designed around a \texttt{Processor} class, which consists of several different components.
An instance of \texttt{Processor} emulates the behaviour of a quantum processor that takes a quantum circuit (\texttt{QubitCircuit}) as well as an initial quantum state (\texttt{qutip.Qobj}) and produces the final state as a (\texttt{qutip.Result}) object. As discussed further below in this section, the key improvements in the new \texttt{qutip-qip} package are the \texttt{Model}, \texttt{GateCompiler}, \texttt{Scheduler} and the \texttt{Noise} classes that allow a modular and flexible design of realistic quantum processors for simulations.

We illustrate our new framework here with an example simulating a 3-qubit Deutsch-Jozsa algorithm on a chain of spin qubits.
We will work through this example and explain briefly the workflow and all the main modules.
We then describe each module in detail in the subsequent subsections.
The simulation of a more complicated circuit, a 10-qubit QFT algorithm, is presented in \cref{sec:qft}.

In \texttt{qutip-qip}, a quantum circuit is represented by an instance of the \texttt{QubitCircuit} class.
The following code defines a circuit of a 3-qubit Deutsch-Jozsa algorithm (see \cref{fig:dj circuit})\footnote{The code examples present in the main text and the Appendices are available at \href{https://www.github.com/boxili/qutip-qip-paper}{github.com/boxili/qutip-qip-paper}. The code is compatible with \texttt{qutip-qip==0.2}.}:
\inputpython{code/code1_main_example.py}{7}{20}

The Deutsch-Jozsa algorithm consists of an oracle constructed using two CNOT gates.
The first two qubits in our circuit take a binary input and will be measured at the end while the last qubit is an ancillary qubit that stores the result of the oracle. The goal is to test if the oracle function is balanced or constant. A constant function returns all $0$ or $1$ for any input, while a balanced function returns $0$ for half of the input combinations and $1$ for the other half.

Among the four different classical inputs ($\{00, 01, 10, 11\}$), for half of them, the oracle returns 0 while for the other half it returns 1. Hence it is a balanced function and, without noise, the measurement of the first two qubits will never be both 0.
One can run the gate-level simulation in the following way:
\inputpython{code/code1_main_example.py}{22}{24}
where we first initialize the state as $\ket{000}$ using \texttt{qutip.basis} and then run the circuit simulation.
By checking the final state, one will see that it has no overlap with $\ket{000}$ or $\ket{001}$. 

The above simulation is at the gate level and is computed by matrix-vector products of the gate operators and the input quantum state. We now describe how to simulate the circuit at the pulse level using \texttt{Processor}.

\subsection{Processor}
\label{sec:processor}
\begin{figure*}[t]
    \centering
    \begin{subfigure}[c]{0.46\textwidth}
    \[
        \begin{array}{c}
        \Qcircuit @C=2em @R=1em {
            \lstick{\ket{0}} & \qw & \gate{H} \barrier{2} & \ctrl{2} & \qw \barrier[-2em]{2} & \gate{H}\\
            \lstick{\ket{0}} & \qw & \gate{H} & \qw & \ctrl{1} & \gate{H}\\
            \lstick{\ket{0}} & \gate{X} & \gate{H} & \targ & \targ & \qw 
            }
        \end{array}
        \]
        \caption{Quantum circuit example}
        \label{fig:dj circuit}
    \end{subfigure}
    \begin{minipage}[c]{0.44\textwidth}
    \begin{tcolorbox}[width=\textwidth]
        {\small
        \makebox[0.8cm]{$\Omega_j^{\alpha}$} Single-qubit rotation around an axis\\ \hspace*{0.8cm} $\alpha=x,y,z$ (colour blue and orange)\\[0.1cm]
        \makebox[0.8cm]{$g_j$} Coupling strength (colour green)\\[0.1cm]
        \makebox[0.8cm]{$\Omega_j^{\textnormal{cr}k}$} The cross-resonance effective \\ \hspace*{0.8cm} interaction (colour green)
        }
    \end{tcolorbox}
    \vspace{0.5cm}
    \end{minipage}
    \\[0.5cm]
    \adjustbox{valign=t,trim=0.3cm 0.3cm}{
    \begin{subfigure}[b]{0.34\textwidth}
        \includegraphics[width=\textwidth]{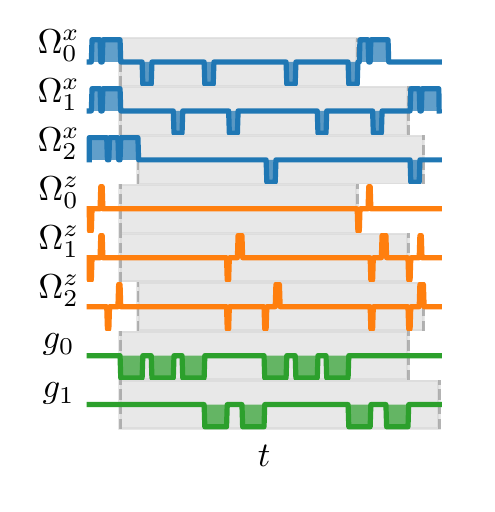}
        \vspace{12pt}
        \caption{Spin chain model}
        \label{fig:DJ spin chain}
    \end{subfigure}
    }
    \adjustbox{valign=t,trim=0.3cm 0.3cm}{
    \begin{subfigure}[b]{0.34\textwidth}
        \includegraphics[width=\textwidth]{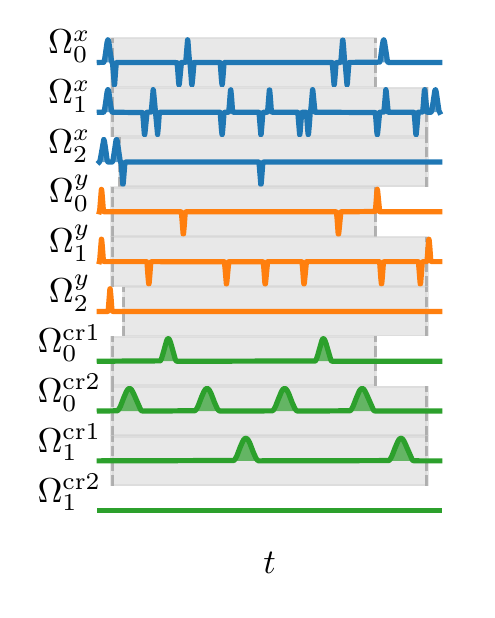}
        \vspace{-25pt}
        \caption{Superconducting qubit model}
        \label{fig:DJ transmon}
    \end{subfigure}
    }
    \adjustbox{valign=t,trim=0.3cm 0.3cm}{
    \begin{subfigure}[b]{0.34\textwidth}
        \includegraphics[width=\textwidth]{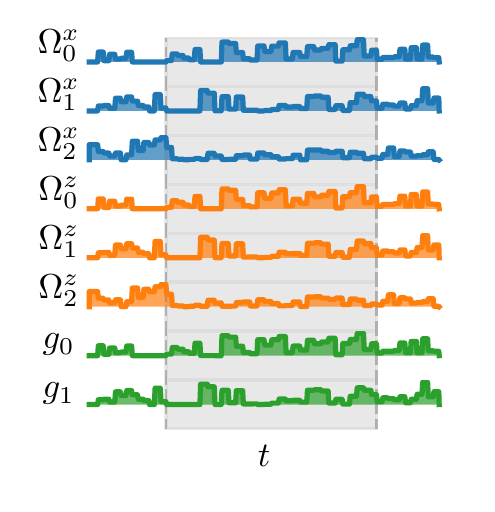}
        \vspace{12pt}
        \caption{Optimal control model}
        \label{fig:DJ optimal control}
    \end{subfigure}
    }
    \vspace{12pt}
    \caption{
    Control pulses generated for a three-qubit Deutsch-Jozsa algorithm (\cref{fig:dj circuit}), where two CNOT gates implement the oracle, which is a balanced function.
    The pulses are compiled using the spin chain model [\cref{eq:ham spin chain}], the superconducting qubits [\cref{eq:superconducting qubits}] and the optimal control algorithm (using GRAPE with the same control Hamiltonian as the spin chain model in \cref{eq:ham spin chain}).
    The symbols for pulse coefficients are defined in the corresponding equations.
    The blue and orange colours denote the two single-qubit control pulses, while green is used for the qubit-qubit interaction.
    For the spin chain and superconducting qubits, the interaction exists only between neighbouring qubits, hence SWAP gates are added to implement the CNOT between the first and third qubits and decomposed into the native gates.
    The grey background marks the pulse duration for the two CNOT gates, where the effect of ASAP scheduling is evident.
    The strength of the compiled pulses, $|c_j(t)|$, is normalized for plotting and should not be compared between different control Hamiltonians.
    Code examples generating these plots are shown in \cref{sec:dj}.
    }
    \label{fig:compiled_pulse}
\end{figure*}

The \texttt{Processor} class handles the routine of a pulse-level simulation.
It first compiles the circuit into a pulse-level description and then simulates the time evolution of the underlying physical system using QuTiP solvers.
For different hardware models and compiling methods the same circuit can be compiled into different pulses, as shown in \cref{fig:DJ spin chain,fig:DJ transmon,fig:DJ optimal control}.
Because of different noise models, the final state also differs from that of the ideal gate-level simulation.

In the following, we choose the spin chain model as an example for the underlying physical system and give an overview of the simulation procedure. We start from initializing a specific type of processor, a subclass of \texttt{Processor} called \texttt{LinearSpinChain}:
\inputpython{code/code1_main_example.py}{25}{26}
where we provide the number of qubits and the $\sigma_x$ drive strength $0.25$MHz.
The other parameters, such as the interaction strength, are set to be the default value.
The decoherence noise can also be added by specifying the coherence times ($T_1$ and $T_2$) which we discuss hereafter.

By initializing this processor with the hardware parameters, a Hamiltonian model for a spin chain system is generated, including the drift and control Hamiltonians ($H_\text{d}$, $H_\text{c}$).
The Hamiltonian model is represented by the \texttt{Model} class and is saved as an attribute of the initialized processor. We provide different predefined models and discuss them more in \cref{sec:model}. In addition, the \texttt{Processor} can also hold simulation configurations such as whether to use a cubic spline interpolation for the pulse coefficients. Such configurations are not directly part of the model but nevertheless could be important for the pulse-level simulation.

Next, we provide the circuit to the processor through the method \texttt{load\_circuit}:
\inputpython{code/code1_main_example.py}{28}{28}
The processor will first decompose the gates in the circuit into native gates that can be implemented directly on the specified hardware model.
Each gate in the circuit is then mapped to the control coefficients and driving Hamiltonians according to the \texttt{GateCompiler} defined for a specific model.
A \texttt{Scheduler} is used to explore the possibility of executing several pulses in parallel. The compiler and scheduler classes will be explained in detail in \cref{sec:compiler,sec:scheduler}.

In addition to the standard compiler, optimal control algorithms in QuTiP can also be used to generate the pulses, which are implemented in \texttt{OptPulseProcessor} (\cref{sec:optimal control}).

With a pulse-level description of the circuit generated and saved in the processor, we can now run the simulation by
\inputpython{code/code1_main_example.py}{29}{32}
The \texttt{run\_state} method first builds a Lindblad model including all the defined noise models (none in this example, but options are discussed below) and then calls a QuTiP solver to simulate the time evolution.
One can pass solver parameters as keyword arguments to the method, e.g., \texttt{tlist} (time sequence for intermediate results), \texttt{e\_ops} (measurement observables) and \texttt{options} (solver options).
In the example above, we record the intermediate state at the time steps given by \texttt{tlist}.
The returned result is a \texttt{qutip.Result} object, which, depending on the solver options, contains the final state, intermediate states and the expectation value.
This allows one to extract all information that the solvers in QuTiP provide.

As for the simulation of noise, simple decoherence noise can be included in the \texttt{Processor} by specifying $T_1, T_2$, e.g.,
\begin{python}
LinearSpinChain(num_qubits=3, t2=30)
\end{python}
More advanced noise models can be represented by the \texttt{Noise} class and added with the method \texttt{Processor.add\_noise}.
The following code is an equivalent way of defining a $T_2$ noise:
\inputpython{code/code1_main_example.py}{34}{35}
In general, the \texttt{Noise} class can be used to represent both decoherence and coherent noise sources.
The former is defined by time-dependent or independent collapse operators and the latter by additional Hamiltonian terms in \cref{eq:Hamiltonian}, with which distortion in the control coefficients or cross-talk can be represented.
In particular, one can define noise that is correlated with the compiled ideal control coefficients through the \texttt{Pulse} class.
They are explained in detail with examples in \cref{sec:noise,sec:pulse}.

Overall, the framework is designed in a modular way so that one can add custom Hamiltonian models, compilers and noise models. We describe in \cref{sec:customization} how this can be done by defining new subclasses.

\subsection{Model}
\label{sec:model}
The pulse-level simulation depends strongly on the modelling of the physical qubits.
In the framework, the physical model is saved as an instance of the \texttt{Model} class in an initialized processor.
A \texttt{Model} object contains the information regarding the specific quantum hardware, including the drift Hamiltonian that cannot be controlled, the available control Hamiltonians and possible noise in the system.
A concrete physical model such as \texttt{SpinChainModel} is defined as a subclass of \texttt{Model}.

For convenience of use, a \texttt{Model} object is automatically generated while initializing a specific \texttt{Processor}, as in the example at the beginning of this section.
To offer more flexibility, \texttt{qutip-qip} provides an equivalent way for the user to define a model and pass it to a \texttt{Processor} object, e.g.,
\inputpython{code/code1_main_example.py}{38}{40}
One can inquire about the properties of a control Hamiltonian through
\inputpython{code/code1_main_example.py}{42}{42}
which returns a tuple consisting of the Hamiltonian as a \texttt{qutip.Qobj} and the indices of the target qubits.
For the predefined models, all available control Hamiltonians can be obtained by
\inputpython{code/code1_main_example.py}{44}{44}
The same interface is also provided in \texttt{Processor} (e.g., \texttt{Processor.get\_control}) for convenience.

In predefined models, these control Hamiltonian terms are simply defined in a dictionary, equivalent to the following code:
\begin{python}
controls = {}
for m in range(num_qubits):
    op = 2 * np.pi * sigmax()
    controls["sx"+str(m)] = (op, m)
\end{python}
which will be accessed by the model object.
Notice that, in general, a model can be correctly recognized by the processor if the method \texttt{Model.get\_control(label)} returns the results in the expected format, regardless of the internal implementation.
For instance, in \cref{sec:customization appendix}, we define it in a different way.
This will be helpful, for instance, in an all-to-all connected system, e.g, using ions or neutral atoms, for which listing all the available combinations of target qubits is tedious.

Models allow one to simulate the physical qubits and their interaction in a more realistic way, e.g., using resonator-induced coupling and including leakage levels.
This is demonstrated by a few predefined models that are implemented as subclasses of \texttt{Model}: the spin chain model, the qubits-resonator model and the fixed-frequency superconducting qubit model. Custom Hamiltonian models can be defined as subclasses as detailed in \cref{sec:customization,sec:customization appendix}. In the following, we illustrate the characteristics of the predefined hardware models in detail.

\subsubsection{Spin Chain model}
The spin-exchange interaction exists in many quantum systems and is one of the earliest types of interaction used in quantum information processing, e.g., in Refs.~\cite{loss1998, Kane1998silicon, buluta2011natural}.
Our predefined \texttt{SpinChainModel} implements a system of a few spin qubits with the exchange interaction arranged in a one-dimensional chain layout with either open ends or closed ends.

The interaction is only possible between adjacent qubits.
For the spin model, the single-qubit control Hamiltonians are $\sigma_j^x$, $\sigma_j^z$, while the interaction is realized by the exchange Hamiltonian $\sigma^x_{j}\sigma^x_{j+1}+
\sigma^y_{j}\sigma^y_{j+1}$.
The control Hamiltonian is given by
\begin{align}
H=&
\sum_{j=0}^{N-1}
\Omega^x_{j}(t) \sigma^x_{j} +
\Omega^z_{j}(t) \sigma^z_{j} \nonumber\\&+ \sum_{j=0}^{N-2}
g_{j}(t)
(\sigma^x_{j}\sigma^x_{j+1}+
\sigma^y_{j}\sigma^y_{j+1}),
\label{eq:ham spin chain}
\end{align}
where $\Omega^x$, $\Omega^y$ and $g$ are the time-dependent control coefficients and $N$ is the number of qubits.

\subsubsection{Qubit-resonator model}
In some experimental implementations, interactions are realized by a quantum bus or a resonator connecting different qubits.
The qubit-resonator model describes a system composed of a single resonator and a few qubits connected to it.
The coupling is kept small so that the resonator is rarely excited but acts only as a mediator for entanglement generation.
The single-qubit control Hamiltonians used are $\sigma_x$ and $\sigma_y$.
The dynamics between the resonator and the qubits is captured by the Tavis-Cummings Hamiltonian,  $\propto\sum_j a^\dagger \sigma_j^{-} + a \sigma_j^{+}$, where $a$, $a^\dagger$ are the destruction and creation operators of the resonator, while $\sigma_j^{-}$, $\sigma_j^{+}$ are those of each qubit.
The control of the qubit-resonator coupling depends on the physical implementation, but in the most general case we have single and multi-qubit control in the form,
\begin{align}
H=
\sum_{j=0}^{N-1}&
\Omega^x_{j}(t) \sigma^x_{j} +
\Omega^y_{j}(t) \sigma^y_{j} + 
g_{j}(t)
(a^\dagger \sigma^{-}_{j} + a \sigma^{+}_{j})\;.
\label{eq:ham cavity qed}
\end{align}
In the numerical simulation, the resonator Hamiltonian is truncated to finite levels. The user can find a predefined \texttt{CavityQEDModel} implementing \cref{eq:ham cavity qed}.

\subsubsection{Superconducting qubit model}
\label{sec:transmon}
Superconducting-circuit qubits have been harnessed to provide artificial atoms for quantum simulation and quantum computing ~\cite{buluta2011natural,Devoret13,Krantz2019,gu2017microwave,Kockum_2019}.
In our model, defined by the \texttt{SCQubitsModel} class, each qubit is simulated by a multi-level Duffing model, in which the qubit subspace is provided by the ground state and the first excited state.
By default, the creation and annihilation operators are truncated at the third level, which can be adjusted, if desired, by the user.
The multi-level representation can capture the leakage of the population out of the qubit subspace during single-qubit gates.
The single-qubit control is generated by two orthogonal quadratures $(a_j^{\dagger}+a_j)$ and $i(a_j^{\dagger}-a_j)$.
Same as the spin chain model, the interaction is possible only between adjacent qubits.
Although this interaction is mediated by a resonator, for simplicity, we replace the complicated dynamics among two superconducting qubits and the resonator with a two-qubit effective Hamiltonian derived in \cite{Magesan2020}.

As an example, we choose the cross resonance interaction in the form of $\sigma^z_{j} \sigma^x_{j+1}$, acting only on the two-qubit levels, which is widely used, e.g., in fixed-frequency superconducting qubits.
We can write the Hamiltonian as
\begin{align}
H = &H_{\textnormal{d}} + 
\sum_{j=0}^{N-1}
\Omega^x_{j} (a_j^{\dagger} + a_j) +
\Omega^y_{j} i(a_j^{\dagger} - a_j)  \nonumber \\
&+\sum_{j=0}^{N-2}
\Omega^{\textnormal{cr}1}_{j} \sigma^z_j  \sigma^x_{j+1} + 
\Omega^{\textnormal{cr}2}_{j} \sigma^x_j \sigma^z_{j+1},
\label{eq:superconducting qubits}
\end{align}
where the drift Hamiltonian $H_{\textnormal{d}}$ is defined by the anharmonicity $\alpha_j$ of the second excited state,
\begin{align}
H_{\textnormal{d}} = \sum_{j=0}^{N-1} \frac{\alpha_j}{2} a_j^{\dagger}a_j^{\dagger}a_j a_j.
\end{align}
The coefficients $\Omega^{\textnormal{cr}1}$ and $\Omega^{\textnormal{cr}2}$ are computed from the qubit-resonator detuning and coupling strength~\cite{Magesan2020}.
With additional single-qubit gates, a CNOT gate can be realized using this type of interaction~\cite{Rigetti2010}.
Using this effective Hamiltonian significantly reduces the size of the Hilbert space in the simulation and allows us to include more qubits.
This flexibility in choosing different levels of detail in the modelling is one of the biggest advantages of this framework, in particular for noise simulation (as illustrated in more detail in \cref{sec:noise}).

\subsection{Compiler}
\label{sec:compiler}

A compiler converts the quantum circuit to the corresponding pulse-level controls $c_j(t)H_j$ on the quantum hardware.
In the framework, it is defined as an instance of the \texttt{GateCompiler} class.
The compilation procedure is achieved through the following steps.

First, each quantum gate is decomposed into the native gates (e.g., rotation over $x$, $y$ axes and the CNOT gate), using the existing decomposition scheme in QuTiP.
If a gate acts on two qubits that are not physically connected, like in the chain model and superconducting qubit model, SWAP gates are added to match the topology before the decomposition. Currently, only 1-dimensional chain structures are supported.

Next, the compiler maps each quantum gate to a pulse-level control description.
It takes the hardware parameter defined in the Hamiltonian model and computes the pulse duration and strength to implement the gate.
For continuous pulses, the pulse shape can also be specified using SciPy window functions (\texttt{scipy.signal.windows}).
A pulse scheduler is then used to explore the possibility of executing multiple quantum gates in parallel, which is explained in detail in \cref{sec:scheduler}.

In the end, the compiler returns a time-dependent pulse coefficient $c_j(t)$ for each control Hamiltonian $H_j$ [see \cref{eq:Hamiltonian}].
They contain the full information to implement the circuit and are saved in the processor.
The coefficient $c_j(t)$ is represented by two NumPy arrays, one for the control amplitude and the other for the time sequence.
For a continuous pulse, a cubic spline is used to approximate the coefficient.
This allows the use of compiled Cython code in QuTiP to achieve better performance.

For the predefined physical models described in the previous subsection, the corresponding compilers are also included and they will be used when calling the method \texttt{Processor.load\_circuit}.
As an example, we compile the three-qubit Deutsch-Jozsa algorithm, shown in \cref{fig:dj circuit}, while the compiled pulses on three different models are plotted in \cref{fig:DJ optimal control,fig:DJ spin chain,fig:DJ transmon}.
From the plots, it is evident that the same circuit is compiled to completely different pulse-level controls:
\begin{itemize}
    \item For the spin chain model (\cref{fig:DJ spin chain}), SWAP gates are added between and after the first CNOT gate, swapping the first two qubits (coefficient $g_0$).
    The SWAP gate is decomposed into three iSWAP gates, while the CNOT is decomposed into two iSWAP gates plus additional single-qubit corrections.
    Both the Hadamard gate and the two-qubit gates need to be decomposed to native gates (iSWAP and rotation on the $x$ and $z$ axes).
    The compiled coefficients are square pulses and the control coefficients on $\sigma_z$ and $\sigma_x$ are also different, resulting in different gate times.
    \item For the superconducting-qubit processor (\cref{fig:DJ transmon}), the compiled pulses have a Gaussian shape.
    This is crucial for superconducting qubits because the second excited level is only slightly detuned from the qubit transition energy.
    A smooth pulse usually prevents leakage to the non-computational subspace.
    Similar to the spin chain, SWAP gates are added to switch the zeroth and first qubit and one SWAP gate is compiled to three CNOT gates.
    The control $\Omega_1^{\textnormal{cr}2}$ [defined in \cref{eq:superconducting qubits}] is not used because there is no CNOT gate that is controlled by the second qubit and acts on the first one.
    \item For the optimal control model (\cref{fig:DJ optimal control}), we use the GRAPE algorithm, where control pulses are piece-wise constant functions.
    We provide the algorithm with the same control Hamiltonian model used for the spin chain model, Eq.~(\ref{eq:ham spin chain}).
    In the compiled optimal signals, all controls are active (non-zero pulse amplitude) during most of the execution time.
    We note that for identical gates on different qubits (e.g., Hadamard), each optimized pulse is different, demonstrating that the optimized solution is not unique, and there are further constraints one could apply, such as adaptions for the specific hardware.
\end{itemize}

As a demonstration of the capability of the simulator, we also compile a 10-qubit QFT algorithm using \texttt{LinearSpinChain}, as shown in \cref{sec:qft}.

To end this subsection, we mention that the gate decomposition is not fully optimized in QuTiP.
Circuit optimization at the level of quantum gates, such as for an optimal number of two-qubit gates, depends on the hardware of interest and is still an open research topic~\cite{Maslov2008,Javadiabhari2015,Haner2018,fosel2021quantum}.
The same holds for mapping the circuit to the topology of the qubits' connectivity~\cite{metodi2006,sargaran2019saqip,Murali2019}.
Because the focus of this simulator is the simulation of the circuit at the physics level, we leave more advanced optimization and scheduling techniques at the gate level for future work.
Instead, we offer the possibility to import quantum circuits defined in other libraries into QuTiP in the QASM format (see \cref{sec:qasm}). This allows possible optimizations elsewhere and then exporting the optimized circuits in QuTiP for a pulse-level simulation.

\subsection{Scheduler}
\label{sec:scheduler}
The scheduling of a circuit consists of an important part of the compilation.
Without it, the gates will be executed one by one and many qubits will be idling during the circuit execution, which increases the execution time and reduces the fidelity.
In the framework, the scheduler is used after the control coefficient of each gate is computed.
It runs a scheduling algorithm to determine the starting time of each gate while keeping the result correct.

The heuristic scheduling algorithm we provide offers two different modes: ASAP (as soon as possible) and ALAP (as late as possible).
In addition, one can choose whether permutation among commuting gates is allowed to achieve a shorter execution time.
The scheduler implemented here does not take the hardware architecture into consideration and assumes that the connectivity in the provided circuit matches with the hardware at this step.

In predefined processors, the scheduler runs automatically when loading a circuit and hence there is no action necessary from the side of the user.
To help explain the scheduling algorithm, we provide here two examples of directly using the \texttt{Scheduler} class.

For gate scheduling, one can use
\inputpython{code/code1_main_example.py}{49}{49}
which, for the 3-qubit Deutsch-Jozsa example (\cref{fig:dj circuit}), returns a list
\begin{python}
[0, 0, 0, 1, 2, 3, 3, 4]
\end{python}
This list denotes the gate cycle of each gate in the circuit.
Here, all gates are assumed to have the same duration.
One can see that that, e.g., the second CNOT and the last Hadamard on the first qubit are grouped together in cycle 3.

For pulse scheduling, one needs to use the \texttt{Instruction} class, which includes information about a specific implementation of a gate on the hardware, e.g., the duration of a gate.
If we assume that all single-qubit gates take a time duration of 1 unit while the CNOT takes a time duration of 2 units, we can rewrite it as
\inputpython{code/code1_main_example.py}{51}{63}
Notice that now we used the ALAP scheduling.
This returns a different list
\begin{python}
[0, 3, 1, 1, 4, 2, 6, 6]
\end{python}
with the starting time of each gate.
In this result, the two CNOT gates (starting time 4 and 2) are exchanged, so that the first Hadamard on the zeroth qubit only needs to start at time step 3.

In the following, we describe our implementation of the pulse scheduler.
The implementation is similar to Ref.~\cite{metodi2006,Guerreschi2018}.
However, we omit the hardware-dependent part but allow gates to have different duration, generalizing it to a pulse scheduler.
We focus on the ASAP scheduling while for the ALAP mode the circuit is reversed before it is passed to the algorithm and then reversed back after the scheduling.

We first represent the dependency among quantum gates in a quantum circuit as a directed acyclic graph.
Each gate is represented by a node and the dependency by arrows.
Gate A is considered dependent on gate B if A has to be executed after B.
This also means that A needs to be executed after all the gates that B depends on.
Hence, there is no loop in the graph.
Next, all gates are divided into different cycles (ignoring the gate duration) according to the dependency graph.
A priority is then assigned to each quantum gate, determined by the time required to execute all the gates that depend on it.
The more time it takes to execute the gates after it, the higher priority is assigned to this gate.
In the end, from the dependency graph and the priority, a list-scheduling algorithm is used to determine the order of the execution and the starting time of each operation.

Unlike scheduling classical gates, a scheduler of quantum gates needs to take the commutation relation into account.
For instance, if two CNOT gates are controlled by the same qubit, but act on two different target qubits, they can be exchanged.
Exploring this flexibility may reduce the total execution time, as shown in the example above.
This is included in the process of building the dependency graph.
All commuting gates are added to the same cycle when computing the priority and the one with the highest priority will be executed first.
In general, more advanced techniques need to be applied to optimize the commuting gates, for instance as discussed in Ref.~\cite{Guerreschi2018}.
However, this becomes more complicated when gates have different execution times.
For simplicity, we omit these advanced techniques in this implementation.

\subsection{Optimal control}
\label{sec:optimal control}
Apart from using compilers with predefined gate-to-pulse maps, one can also use the optimal control algorithm in QuTiP to find optimized control pulses.
The algorithm can take arbitrary control Hamiltonians as input and uses quantum control function optimisation, based on open-loop quantum control theory~\cite{DAlessandro2008_Quantum_Control} to find the best pulses.
For a set of given control Hamiltonians $H_j$, the optimal control module uses classical algorithms to optimize the control function $c_j(t)$ in \cref{eq:Hamiltonian}.
Parameters of control pulses for realizing individual gates, sequences and hence complete circuits, are generated automatically through multi-variable optimization targeting maximum fidelity with the evolution described by the circuit.

The optimal control module in QuTiP supports both the GRAPE \cite{Khaneja2005,Machnes2010_DYNAMO} and the CRAB algorithms \cite{Caneva2011,Doria2011_CRAB_PRL}.
The interface to use these algorithms in \texttt{qutip-qip} is implemented via the \texttt{OptPulseProcessor} class. 
One first provides the available control Hamiltonians that characterize the physical controls on the system, which, e.g., can be provided as an instance of the \texttt{Model} class, such as the \texttt{SpinChainModel}.
Upon loading the quantum circuit, each quantum gate is expanded to a unitary acting on the full Hilbert space and passed to the optimal control algorithm as the desired target.
The returned pulses that drive this are concatenated to complete a full circuit simulation of the physical control sequences.
An example of optimized pulses is shown in \cref{fig:DJ optimal control} and the code can be found in \cref{sec:dj}.

\subsection{Noise}
\label{sec:noise}
The noise module allows one to add control and decoherence noise following the Lindblad description of open quantum systems [\cref{Eq:Lindblad}].
Compared to the gate-based simulator (\cref{sec:circuit}), this provides a more practical and straightforward way to describe the noise.
In the current framework, noise can be added at different layers of the simulation, allowing one to focus on the dynamics of the dominant noise, while representing other noise, such as single-qubit relaxation, as collapse operators for efficiency.
Depending on the problem studied, one can devote the computing resources to the most relevant type of noise.

Apart from imperfections in the Hamiltonian model and circuit compilation, the \texttt{Noise} class in the current framework defines deviations of the real physical dynamics from the compiled one.
It takes the compiled pulse-level description of the circuit (see also \cref{sec:pulse}) and adds noise elements to it, which allows defining noise that is correlated to the compiled pulses.
In the following, we detail the three different noise models already available in the current framework.

\paragraph{Noise in the hardware model.}
The Hamiltonian model defined in the \texttt{Model} class may contain intrinsic imperfections of the system and hence the compiled ideal pulse does not implement the ideal unitary gate.
Therefore, building a realistic Hamiltonian model usually already introduces noise to the simulation.
An example is the superconducting-qubit processor model (\cref{sec:transmon}), where the physical qubit is represented by a multi-level system.
Since the second excitation level is only weakly detuned from the qubit transition frequency, the population may leak out of the qubit subspace.
Another example is an always-on ZZ type cross-talk induced by interaction with higher levels of the physical qubits~\cite{Mundada2019}, which is also implemented for the superconducting qubit model.

\paragraph{Control noise.}
The control noise, as the name suggests, arises from imperfect control of the quantum system, such as distortion in the pulse amplitude or frequency drift.
The simplest example is the random amplitude noise on the control coefficient $c_j(t)$ in \cref{eq:Hamiltonian}.

As a demonstration of control noise, we simulate classical cross-talk-induced decoherence between two neighbouring ion trap qubits described in \cite{piltz2014trapped}.
We build a two-qubit \texttt{Processor}, where the second qubit is detuned from the first one by $\delta=1.852$~MHz.
A sequence of $\pi$-pulses with Rabi frequency of $\Omega = 20$~KHz and random phases are applied to the first qubit.
We define noise such that the same pulse also applies to the second qubit.
Because of the detuning, this pulse does not flip the second qubit but subjects it to a diffusive behaviour, so that the average fidelity of the second qubit with respect to the initial state decreases.
This decreasing fidelity is shown experimentally in Figure 3a of Ref.~\cite{piltz2014trapped}.

\begin{figure}
\centering
\includegraphics[width=0.95\linewidth]{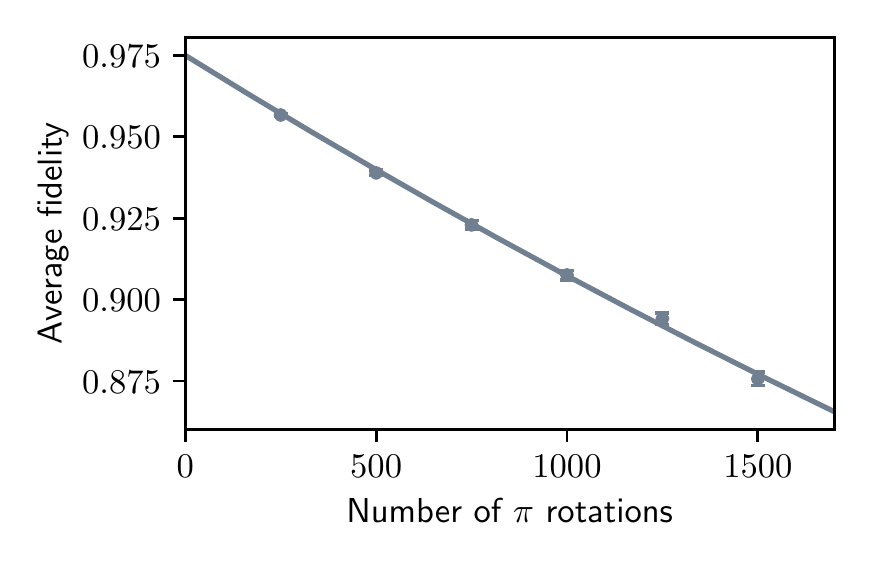}
\caption{An example of simulated classical cross-talk-induced decoherence between neighbouring qubits in an ion trap system.
The randomized benchmarking protocol is adopted from Piltz{\it~et~al.}~\cite{piltz2014trapped} and the figure reproduces the measured fidelity decay in figure 3a of that work. We build a custom \texttt{Processor} and \texttt{Noise} object to define classical cross-talk noise and perform our simulations.
It shows the average fidelity of the qubit when a sequence of single-qubit $\pi$ rotations with random phase is applied to its direct neighbour.
The cross-talk is simulated by adding control pulses to the neighbouring qubits with a strength proportional to that of the target qubit and detuned by the difference of the qubit transition frequency.
Each point is sampled from 1600 repetitions. We set the detuning $\delta=1.852$~MHz, the Rabi frequency $\Omega = 20$~KHz and the cross-talk ratio $\lambda=1$.
}
\label{fig:cross-talk}
\end{figure}

Here, we reproduce these results with our two-qubit \texttt{Processor} in \cref{fig:cross-talk}.
We start with an initial state of fidelity 0.975 and simulate the Hamiltonian
\begin{align}
H=\Omega(t)(\sigma^x_0 + \lambda \sigma^x_1) + \delta\sigma^z_1
,
\label{eq:hamcrosstalk}
\end{align}
where $\lambda$ is the ratio between the cross-talk pulse's amplitudes.
The plot in Figure~\ref{fig:cross-talk} shows a similar fidelity decay curve as the experimental result, but includes only the contribution of cross-talk, while in the experimental result other noise sources may exist.
This kind of simulation provides a way to identify noise contributions from different sources.
The code is described in detail in \cref{sec:customization appendix}, as an example of a custom noise model.

\paragraph{Lindblad noise.}
The Lindblad noise originates from the coupling of the quantum system with the environment (e.g.,~a thermal bath) and leads to loss of information.
It is simulated by collapse operators and results in non-unitary dynamics~\cite{breuer2002theory,Lidar2019}.

The most commonly used type of Lindblad noise is decoherence, characterized by the coherence time $T_1$ and $T_2$ (dephasing).
For the sake of convenience, one only needs to provide the parameter \texttt{t1}, \texttt{t2} to the processor and the corresponding operators will be generated automatically.
Both can be either a number that specifies one coherence time for all qubits or a list of numbers, each corresponding to one qubit.

For $T_1$, the operator is defined as $a/\sqrt{T_1} $ with $a$ as the destruction operator.
For $T_2$, the operator is defined as $a^{\dagger}a\sqrt{2/T_2^*} $, where $T_2^*$ is the pure dephasing time given by $1/T_2^*=1/T_2-1/(2T_1)$.
In the case of qubits, i.e., a two-level system, the destruction operator $a$ is truncated to a two-level operator and is consistent with \cref{Eq:Lindblad}.
Constant $T_1$ and $T_2$ can be provided directly when initializing the \texttt{Processor}.
Custom collapse operators, including time-dependent ones, can be defined through \texttt{DecoherenceNoise}.
For instance, the following code defines a collapse operator using \texttt{qutip.sigmam()} and increases linearly as time:
\begin{python}
tlist = np.linspace(0, 30., 100)
coeff = tlist * 0.01
noise = DecoherenceNoise(
    sigmam(), targets=0,
    coeff=coeff, tlist=tlist)
proc.add_noise(noise)
\end{python}
Similar to the control noise, the Lindblad noise can also depend on the control coefficient.

\begin{figure}
    \centering
    \begin{tikzpicture}
        \node[anchor=south west] at (0,0) {\includegraphics[width=0.5\textwidth]{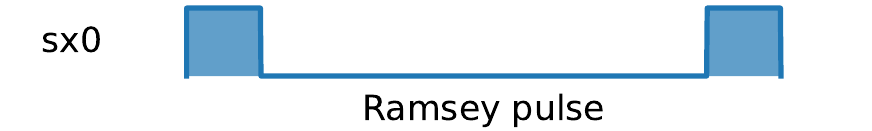}};
        \node[] at (2.65, 1)   (a) {};
        \node[] at (7.05, 1)    (b) {};
        \draw[<->, line width=1pt] (a)--(b);
        \node[] at (4.85, 1.3)    (b) {$t$};
        \node[] at (2.3, 0.3)    (b) {$t_{\pi/2}$};
        \node[] at (7.45, 0.3)    (b) {$t_{\pi/2}$};
    \end{tikzpicture}
    \includegraphics[width=0.5\textwidth]{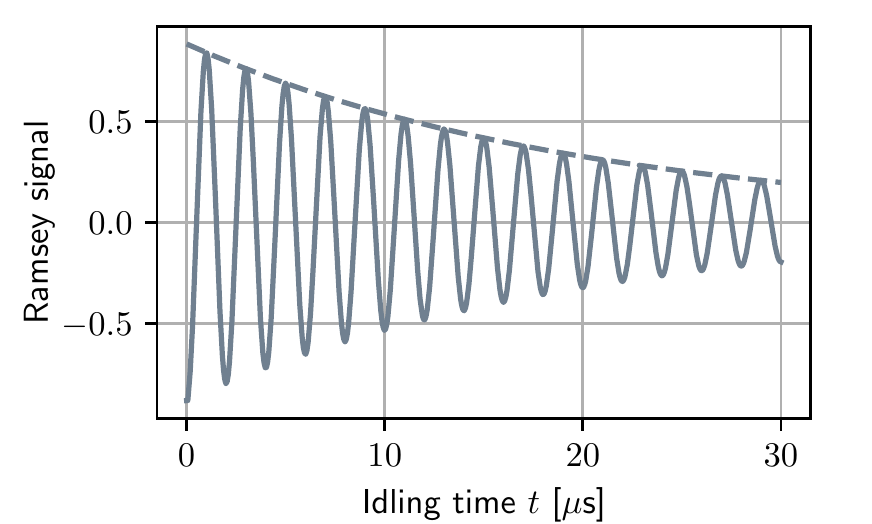}
    \caption{The Ramsey pulse and the simulated measurement results.
    The quantum system is subjected to a rotation around the $z$ axis and a $T_2$ decoherence.
    The Ramsey pulse consists of two $\pi/2$ rotations separated by an idling time $t$.
    The expectation value of the measurement for different idling time is recorded.
    The solid line represents the measured expectation value.
    The dashed line is the fitted exponential decay.
    Due to the imperfect preparation of the superposed state, the envelope does not start from one.
    }
    \label{fig:decoherence}
\end{figure}
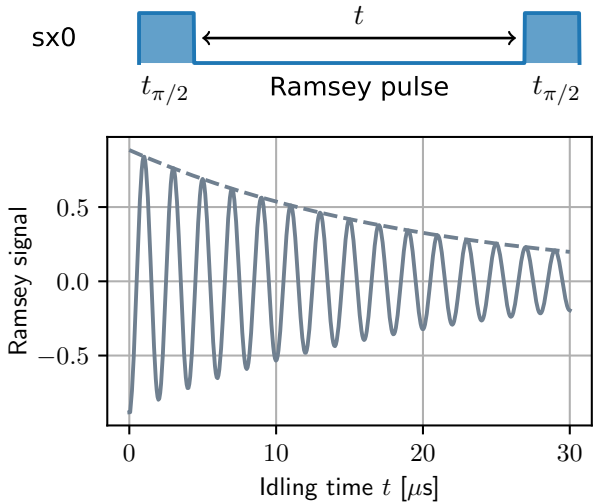

In order to demonstrate the simulation of decoherence noise, we build an example that simulates a Ramsey experiment as a quantum circuit run on a noisy \texttt{Processor}.
The Ramsey experiment consists of a qubit that is initialized in the excited state, undergoes a $\pi/2$ rotation around the $x$ axis, idles for a time $t$, and is finally measured after another $\pi/2$ rotation:

\inputpython{code/code4_decoherence.py}{20}{45}
In the above block, we use the linear spin chain processor just for its compiler and do not use any of its default Hamiltonians.
Instead, we define an always-on drift Hamiltonian $\sigma^z$ with frequency $f=0.5$~MHz, an on-resonant $\sigma^x$ drive with an amplitude of $0.1/2$~MHz and the coherence time $T_2=10/f$.
For different idling time $t$, we record the expectation value with respect to the observable $\sigma^z$, which is plotted in \cref{fig:decoherence} as the solid curve.
As expected, the envelope follows an exponential decay characterized by $T_2$ (dashed curve).
Notice that, because $\pi/2$-pulses are simulated as a physical process, the fitted decay does not start from 1.
This demonstrates a way to include state preparation error into the simulation.

\subsection{Pulse}
\label{sec:pulse}
As discussed before, in this simulation framework, we compile the circuit into pulse-level controls $c_j(t) H_j$ [\cref{eq:Hamiltonian}] and simulate the time evolution of the physical qubits.
In this subsection, we describe how the dynamics is represented internally in the workflow of \texttt{qutip-qip}, which is useful for understanding the simulation process as well as defining custom pulse-dependent noise.

A control pulse, together with the noise associated with it, is represented by a class instance of \texttt{Pulse}.
When an ideal control is compiled and returned to the processor, it is saved as an initialized \texttt{Pulse} object, equivalent to the following code:
\begin{python}
coeff = np.array([1.])
tlist = np.array([0., np.pi])
pulse = Pulse(
    sigmax()/2, targets=0, tlist=tlist,
    coeff=coeff, label="pi-pulse")
\end{python}
This code defines a $\pi$-pulse implemented using the term $\sigma_x$ in the Hamiltonian that flips the zeroth qubit specified by the argument \texttt{targets}. The pulse needs to be applied for the duration $\pi$ specified by the variable \texttt{tlist}. The parameters \texttt{coeff} and \texttt{tlist} together describe the control coefficient $c(t)$.
Together with the provided Hamiltonian and target qubits, an instance of \texttt{Pulse} determines the dynamics of one control term.

With a \texttt{Pulse} initialized with the ideal control, one can define several types of noise, including the Lindblad or control noise as described in \cref{sec:noise}.
An example of adding a noisy Hamiltonian as control noise through the method \texttt{add\_control\_noise} is given below:
\begin{python}
pulse.add_control_noise(
    sigmaz(), targets=[0], tlist=tlist,
    coeff=coeff * 0.05)
\end{python}
The above code snippet adds a Hamiltonian term $\sigma_z$, which can, for instance, be interpreted as a frequency drift.
Similarly, collapse operators depending on a specific control pulse can be added by the method \texttt{Pulse.add\_lindblad\_noise}.

In addition to a constant pulse, the control pulse and noise can also be provided as continuous functions. In this case, both \texttt{tlist} and \texttt{coeff} are given as NumPy arrays and a cubic spline is used to interpolate the continuous pulse coefficient.
This allows using the compiled Cython version of the QuTiP solvers that have a much better performance than using a Python function for the coefficient.
The option is provided as a keyword argument \texttt{spline\_kind="cubic"} when initializing \texttt{Pulse}.
Similarly, the interpolation method can also be defined for \texttt{Processor} using the same signature.

\subsection{Adding custom hardware models}
\label{sec:customization}

As it is impractical to include every physical platform, we provide an interface that allows one to customize the simulators.
In particular, the modular architecture allows one to conveniently overwrite existing modules for customization.

To define a customized hardware model, the minimal requirements are a set of available control Hamiltonians $H_j$, and a compiler, i.e., the mapping between native gates and control coefficients $c_j$.
One can either modify an existing subclass or write one from scratch by creating a subclass of the two parent classes \texttt{Model} and \texttt{GateCompiler}.
Since different subclasses share the same interface, different models and compilers can also be combined to build new processors.

Moreover, this customization is not limited to Hamiltonian models and compiler routines.
In principle, measurement can be defined as a customized quantum gate and the measurement statistics can be extracted from the obtained density matrix.
A new type of noise can also be implemented by defining a new \texttt{Noise} subclass, which takes the compiled ideal \texttt{Pulse} and adds noisy dynamics on top of it.

An example of building a customized \texttt{Model} and \texttt{GateCompiler}, with custom types of noise, is provided in \cref{sec:customization appendix}.

\section{Importing and exporting circuits in QASM format}
\label{sec:qasm}
As pointed out in \cref{sec:compiler}, it is impractical to include all the advanced techniques for circuit optimization and scheduling.
To allow integration with other packages, we support import and export of circuits in the intermediate Quantum Assembly Language (QASM) format~\cite{cross2017open}. While there are different intermediate representations for quantum programs, and more specifically quantum circuits, including cQASM \cite{khammassi2018cqasm}, \texttt{qutip-qip} provides support for OpenQASM. OpenQASM is an imperative programming language that can be used to describe quantum circuits in a back-end agnostic manner. 

QuTiP includes a module to import and export quantum circuits compatible with the OpenQASM 2.0 standard \cite{cross2017open}. OpenQASM 2.0 allows concise quantum circuit definitions including useful features like custom unitaries and defining groups of qubits over which a common gate can be applied simultaneously. Due to compatibility with multiple libraries such as Qiskit and Cirq, it is an easy way to transfer quantum circuits between these libraries and \texttt{qutip-qip}.

As an example, we use again the $3$-qubit Deutsch-Jozsa circuit (\cref{fig:dj circuit}).
The following block defines the same circuit in the QASM format:
\inputpython{code/code7_deutsch_jozsa.qasm}{0}{14}
It can be saved as a \texttt{.qasm} file (such as \texttt{``deutsch-jozsa.qasm''} in our example below).

Every QASM file imported to \texttt{qutip-qip} requires the two header statements at the beginning of the file. The line \texttt{OPENQASM 2.0} declares that the file adheres to the OpenQASM 2.0 standard. The keyword \texttt{include} processes a file that contains definitions of some QASM gates. It is available in the OpenQASM repository (as a standard file) and is included with the QASM file exported by \texttt{qutip-qip} (and also by Qiskit/Cirq). This circuit can be easily imported into \texttt{qutip-qip} using the \texttt{read\_qasm} method in the following manner:
\inputpython{code/code5_deutsch_jozsa-qasm.py}{1}{2}

Furthermore, using the \texttt{strmode} option for \texttt{read\_qasm} function, we can import the circuit described in a string object. Once a quantum circuit is defined, we can also export it to the QASM format and save it as a file using the \texttt{save\_qasm} method: 
\inputpython{code/code5_deutsch_jozsa-qasm.py}{4}{5}
The circuit can then be simulated with other packages. It is also possible to output the circuit as a string using \texttt{circuit\_to\_qasm\_str} or print it out using \texttt{print\_qasm}.

\section{Conclusion}
\label{sec:conclusion}
In this work, we presented a framework for pulse-level quantum circuit simulation that can be used to study noisy quantum devices simulated on classical computers.
This framework builds on existing solvers and the quantum circuit model offered by QuTiP. We expanded the noise modeling capabilities with ad-hoc features for the simulation of controls in noisy quantum circuits, such as providing the option to inject coherent noise in pulses.

We provided a few predefined quantum hardware models, compiling and scheduling routines, as well as noise models, which can be adjusted to devote limited computing resources to the most relevant physical dynamics during the study of noise. We showed the simulation capabilities by illustrating how results obtained on cross-talk noise characterization for an ion-trap-based quantum processor can be easily replicated with this toolbox. Moreover, we provided an example of the simulation of Lindblad noise for a Ramsey experiment.  

Due to the modular design, the framework introduced here can be integrated with more hardware models, gate decomposition and optimization schemes. In particular, the simulation of processors supporting bosonic models for quantum information processing, including quantum error correction schemes, is especially suitable within the current framework. 
Represented as customized gates, state preparation and measurement can also be simulated as a noisy physical process. 

Pulse-level simulation could be helpful in quick verification of experimental results, developing quantum algorithms, such as variational quantum algorithms~\cite{Alam2020,haug2021optimal, bharti2021noisy,Magann2021}, and testing compiling and scheduling schemes~\cite{Murali2019} with realistic noise models \cite{erik_2021_pygsti}.
Through hardware simulation and noise simulation, quantum error correction code and quantum mitigation protocols can also be studied, for example, simulating pulse-level and digital zero-noise extrapolation \cite{Kandala2019,Giurgica_Tiron_2020,larose2020mitiq}. 

Moreover, the noise characterization in model devices \cite{Dahlhauser21} and the impact of non-Markovian types of noise could be further evaluated \cite{schultz2020schwarma,lambert2020bofinheom}.  Future development in QuTiP aims at providing a unified interface to the open system solvers, which would enable a simpler integration with \texttt{qutip-qip}. 
This approach also has a potential to be integrated with other quantum control software such as \texttt{qupulse} ~\cite{simon_humpohl_2020} and \texttt{C3} \cite{Wittler_2021}. In particular, the features here introduced may be a useful tool to investigate from a novel perspective many-body dynamical properties of quantum circuits, such as for measurement-induced phase transitions \cite{PhysRevX.9.031009}, chaotic dynamics and information scrambling \cite{Blok21}.   

Planned developments in \texttt{qutip} and \texttt{qutip-qip} will enable the use of alternative quantum control optimization algorithms, that is options other than the GRAPE and CRAB algorithms that are currently supported. Most immediately Krotov-type algorithm support could be added through integration with \texttt{qucontrol-krotov} \cite{Goerz_2019_SciPost}, which is already closely aligned with QuTiP. Further opportunities for development and integration with the main QuTiP package include the development of an implementation of the GOAT algorithm \cite{Machnes2018_GOAT}, in which \texttt{qutip}'s solvers of various kinds can be used effectively. This could then also be available for optimization of circuit controls to simulate universal gate operations \cite{Dong_2015,Dong_2016}.

Another direction of development is the integration with other software frameworks, in the ecosystem of quantum open source software, where considerable duplication exists. Even with respect to the quantum intermediate representation of quantum circuits, standards are not yet solidified. For example, we have connected \texttt{qutip-qip} with OpenQASM 2.0, thus providing an access point to any major framework. More sophisticated features are expected in the upcoming OpenQASM 3.0 standard \cite{cross2021openqasm}, including classical computation specifications and the option for pulse-level definitions for gates. Extending QuTiP support to OpenQASM 3.0 will be an important step in cross-package compatibility with respect to pulse-level quantum circuit simulation and their integration with real hardware. 

The use of \texttt{qutip-qip} for open quantum hardware is an especially intriguing direction of research and development. One could envision this framework as the backbone for API inter-connectivity between simulation and hardware control in research labs with different technologies \cite{simon_humpohl_2020}.
\section*{Acknowledgements}
We would like to thank the whole community of contributors and users of the QuTiP project and in particular those who developed the quantum circuit representation in the \texttt{qutip.qip} module \cite{qutipqip} --  in particular Anubhav Vardhan, Robert Johansson, and Paul Nation. We also thank Jake Lishman, \`{E}ric Gigu\`{e}re, Purva Thakre and Simon Cross for work on the QuTiP project. We thank Andrea Mari, Anton Frisk Kockum, Daniel Burgarth, Sebastian Grijalva for useful discussions and comments on the manuscript. Part of this code development has been supported by NumFOCUS and Google Summer of Code. F.~N.~is supported in part by
Nippon Telegraph and Telephone Corporation (NTT) Research, 
the Japan Science and Technology Agency (JST) [via 
the Quantum Leap Flagship Program (Q-LEAP), 
the Moonshot R\&D Grant Number JPMJMS2061, and 
the Centers of Research Excellence in Science and Technology (CREST) Grant No.~JPMJCR1676], 
the Japan Society for the Promotion of Science (JSPS) 
[via the Grants-in-Aid for Scientific Research (KAKENHI) Grant No.~JP20H00134 and the 
JSPS–RFBR Grant No.~JPJSBP120194828],
the Army Research Office (ARO) (Grant No.~W911NF-18-1-0358),
the Asian Office of Aerospace Research and Development (AOARD) (via Grant No.~FA2386-20-1-4069), and 
the Foundational Questions Institute Fund (FQXi) via Grant No.~FQXi-IAF19-06.
N.~S.~acknowledges partial support from the
U.S. Department of Energy, Office of Science, Office of
Advanced Scientific Computing Research, Accelerated
Research in Quantum Computing under Award Number de-sc0020266. S.~A.~acknowledges support from the Knut and Alice Wallenberg Foundation through the Wallenberg Centre for Quantum Technology (WACQT).
\section*{Data availability}
The code examples present in the main text and the Appendices are available at \href{https://www.github.com/boxili/qutip-qip-paper}{github.com/boxili/qutip-qip-paper}. A version compatible with the latest distribution of \texttt{qutip-qip} can be found at \href{https://www.github.com/qutip/qutip-qip}{github.com/qutip/qutip-qip}.

\clearpage

\onecolumn
\appendix

\onecolumn
\section{Simulating the Deutsch-Jozsa algorithm}
\label{sec:dj}
In this section, we show the code example of simulating the 3-qubit Deutsch-Jozsa algorithm on three different hardware models: the spin chain model, the superconducting qubits, and the optimal control model:

\inputpython{code/code2_dj_algorithm.py}{17}{59}

In the above code block, we first define the Deutsch-Jozsa algorithm, same as the circuit shown in \cref{fig:compiled_pulse}.
We then run the circuit on various hardware models.
For the spin model and superconducting qubits, a Hamiltonian model and a compiler are already predefined and one only needs to load the circuit and run the simulation.
Hardware parameters, such as the $T_1$ and $T_2$ times, qubit frequencies and coupling strength, can be given as parameters to initialize the processor.
For optimal control, we use the control Hamiltonians of the spin chain model and provide a few parameters for the optimization routine in QuTiP, such as the maximal pulse amplitude and the number of time slots for each gate.
For details, please refer to the QuTiP documentation ({\url{http://qutip.org/docs/latest/index.html}}).

The generated control pulses are shown in \cref{fig:compiled_pulse} and can be obtained by the method:
\begin{python}
processor.plot_pulses()
\end{python}

\begin{figure}
    \centering
    \begin{subfigure}[b]{0.45\linewidth}
    \centering
    \includegraphics[width=\textwidth]{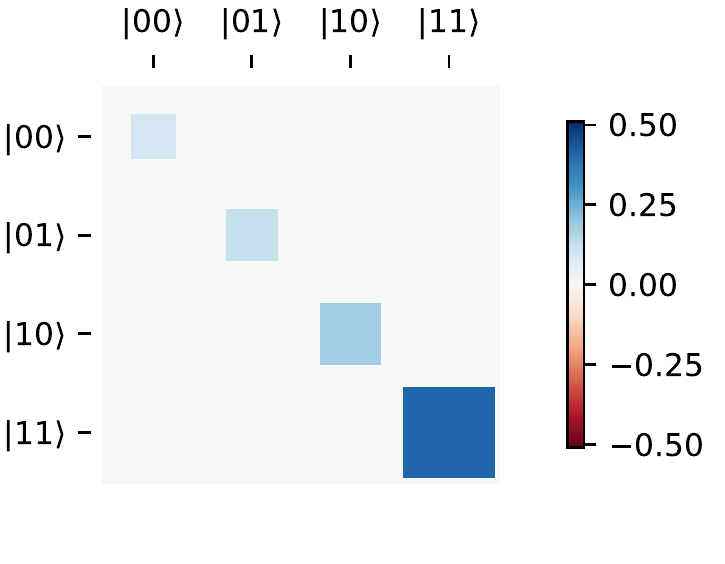}
    \caption{}
    \label{fig:hinton}
    \end{subfigure}
    \begin{subfigure}[b]{0.49\linewidth}
    \centering
    \includegraphics[width=\textwidth]{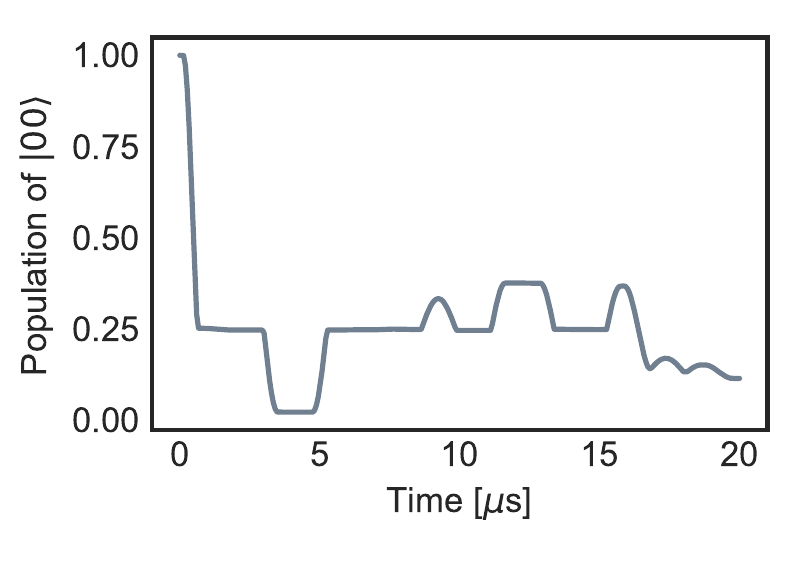}
    \caption{}
    \label{fig:population}
    \end{subfigure}
    \caption{The Hinton diagram of the final density matrix (\cref{fig:hinton}) and the population of the $\ket{00}$ state during the circuit execution (\cref{fig:population}) for the first two qubits in the circuit shown in \cref{fig:compiled_pulse}. The Hinton diagram is a visual representation of the complex-valued density matrix. The shade and size of the blocks are determined by the absolute value of the density matrix element and the color blue (red) denotes whether the real part of the density matrix is positive (negative). For an ideal Deutsch-Jozsa algorithm with a balanced oracle. The first two qubits should end up having no overlap with the ground state. This is not exactly the case in the plot because we define a finite $T_2$ time.}
    \label{fig:dj result}
\end{figure}

Because we are doing a simulation, we have access both to the final states as a density matrix and the information of the states during the evolution.
We demonstrate this in \cref{fig:dj result}.
By construction, the measured result of the first two qubits of a perfect Deutsch-Jozsa algorithm with a balanced oracle should not overlap with the state $\ket{00}$.
This agrees with the small population of the state $\ket{00}$ in the Hinton diagram (\cref{fig:hinton}).
The population is not exactly zero because we define a $T_2$ decoherence noise.
In addition, we can also extract information during the circuit execution, e.g., the population as a function of time (\cref{fig:population}).

\begin{figure*}
    \centering
    \includegraphics[width=6.6in]{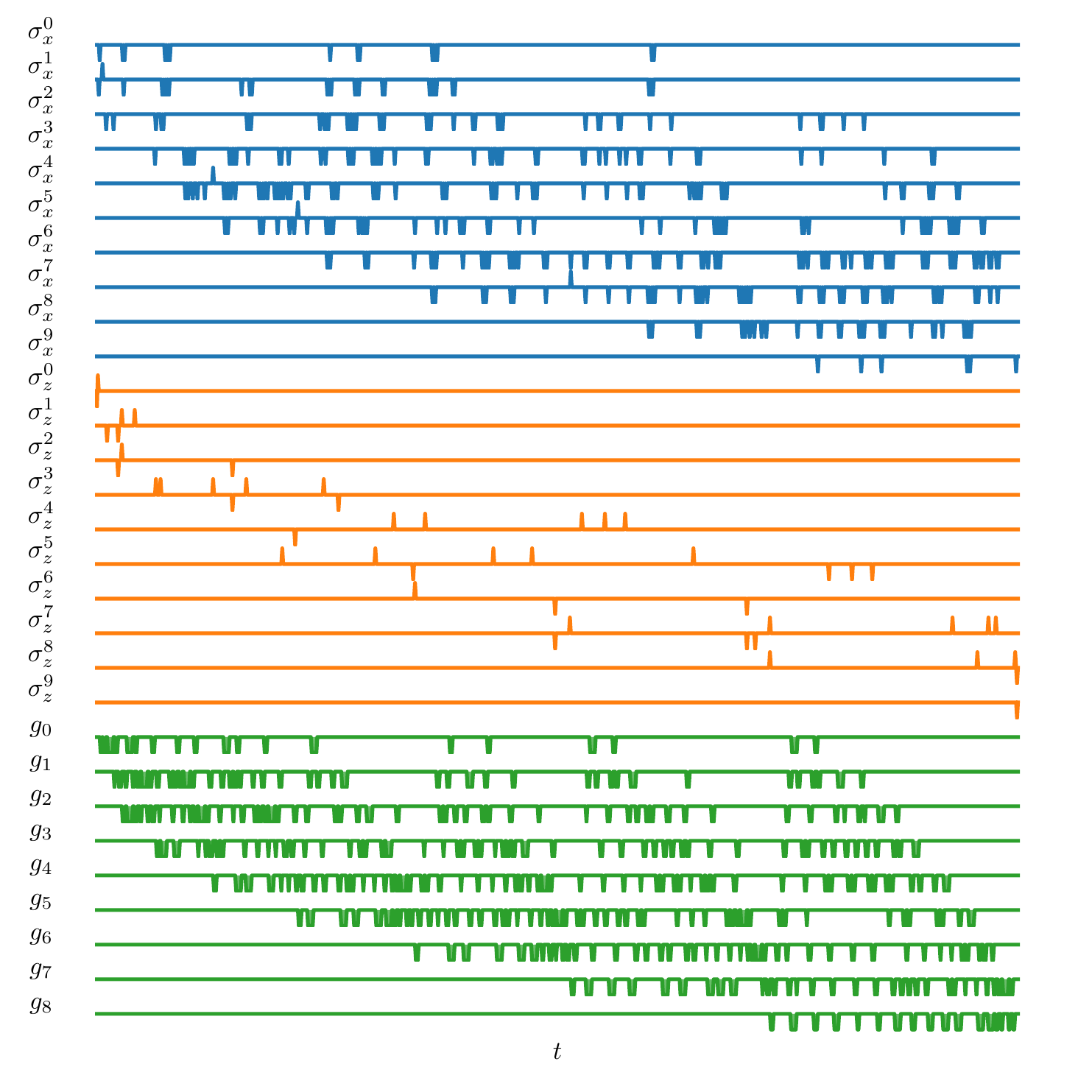}
    \includegraphics[width=5.5in]{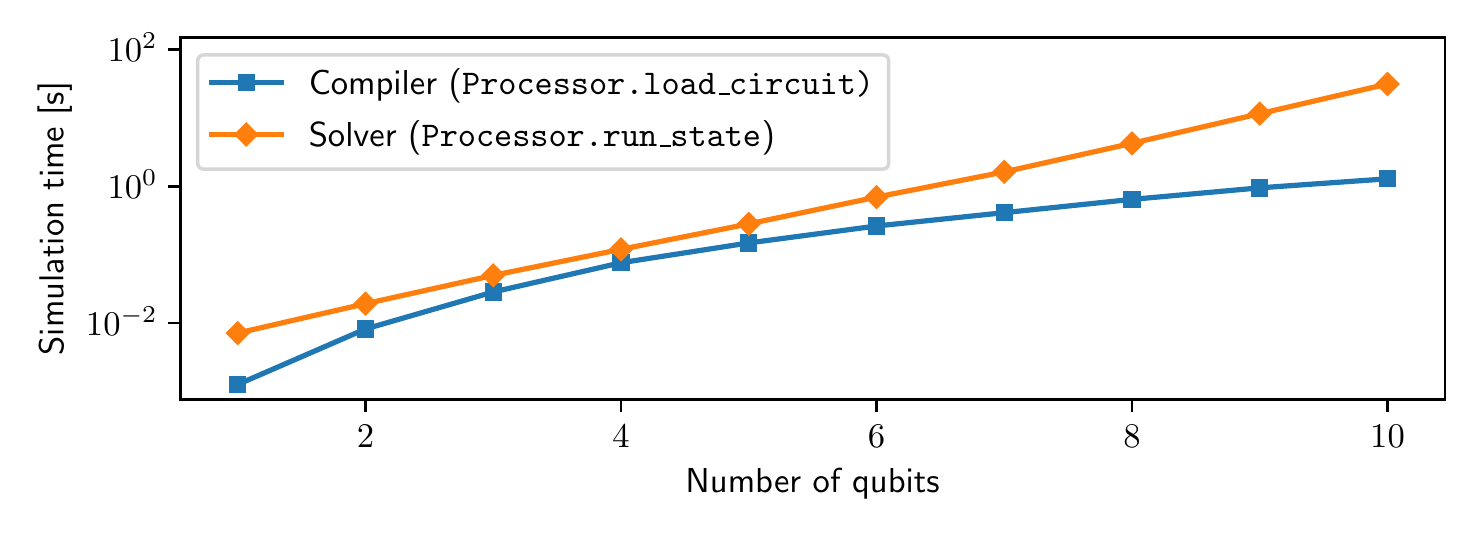}
    \caption{{\bf Top:} Compiled pulses for a 10-qubit QFT circuit using the linear spin chain model (see \cref{fig:DJ spin chain} and \cref{sec:model}).
    The colors and notation used are the same as in \cref{fig:compiled_pulse}.
    The blue and orange colours denote the single-qubit control while the green colour the exchange interactions.
    {\bf Bottom:}
    Simulation time of the QFT algorithm using the spin chain model as a function of the number of qubits, $N=1,2,...,10$, on a commercial CPU with a single thread.
    We plot both the compilation time (\texttt{Processor.load\_circuit}) and the time used to solve the dynamics (\texttt{Processor.run\_state}).
    }
    \label{fig:qft pulse}
\end{figure*}

\section{Compiling and simulating a 10-qubit Quantum Fourier Transform (QFT)}
\label{sec:qft}
In this section, we simulate a 10-qubit Quantum Fourier Transform (QFT) algorithm.
The QFT algorithm is one of the most important quantum algorithms in quantum computing~\cite{nielsen2002quantum}.
It is, for instance, part of the Shor algorithm for integer factorization.
The following code defines a 10-qubit QFT algorithm using CNOT and single qubit rotations and runs the simulation both at the gate level and at the pulse level.

\inputpython{code/code8_qft.py}{13}{28}

We plot the compiled pulses and perform a study of the simulation time in the top and bottom panels of \cref{fig:qft pulse}, respectively.
The top panel of \cref{fig:qft pulse} shows the control pulses $\sigma_x^i$ (blue curves), $\sigma_y^i$ (orange curves) and $g_i$ (green curves) for the spin chain model processor (\cref{sec:model}), where $i=0,...,9$ counts the qubits.
The pulses plotted implement the QFT algorithm represented in the native gates of the spin chain model, with single-qubit gates marked by rotations over the $x$- and $z$-axes and the iSWAP gate implemented through the spin-spin exchange interaction, marked by $g_i$.
While the sign for single-qubit drive denotes the phase of the control pulse, the negative sign in the coupling strengths $g_i$ is only a result of the convention used in the definition of the interaction, defined in \cref{eq:ham spin chain}.

In the bottom panel of \cref{fig:qft pulse}, we study the time it takes to simulate the dynamics for the QFT algorithm on the spin chain processor, from 1 to 10 qubits. 
We divide the simulation between compilation and solution of the dynamical equation. 
The compilation of the algorithm (blue squares in the bottom panel of \cref{fig:qft pulse}) includes native-gate gate decomposition, scheduling, and mapping to control pulses (as shown in the top panel). For 10 qubits, the compilation takes about one second, whereas the overall simulation time takes about half a minute on a commercial CPU (Intel i7 8700 with Max Turbo Frequency 4.60 GHz) with a single thread.
Indeed, the overall simulation time is dominated by the task of solving the Schr\"odinger equation: this increases linearly with the circuit depth and exponentially with the size of the Hilbert space (orange diamonds in the bottom panel of \cref{fig:qft pulse}).
The proportion of time used for the compilation with respect to the total simulation time decreases as the number of qubits in the QFT algorithm grows. As expected, we find that the bottleneck for the simulation of larger processors lies in the solution of the dynamics.

Note that, because of the pulse-level nature of the simulation, the overall simulation time also depends on the typical frequency characterizing the dynamics.
In the above simulation, the maximum frequency in the Hamiltonian is about 1 MHz while the time scale of the quantum circuit is about 2 ms.
No collapse operators are included.
The simulation time may increase if decay or high-frequency coherent noise are included.

\section{Customizing the physical model and noise}
\label{sec:customization appendix}
In the following, we show a minimal example of constructing Hamiltonian models and compilers:

\inputpython{code/code3_customize.py}{15}{131}

In this example, we first build a Hamiltonian model called \texttt{MyModel}.
For simplicity, we only include two single-qubit control Hamiltonians: $\sigma_x$ and $\sigma_y$.
We then define the compiling routines for the two types of rotation gates RX and RY.
In addition, we also define a rotation gate with mixed X and Y quadrature, parameterized by a phase $\phi$, $\cos(\phi)\sigma_x+\sin(\phi)\sigma_y$.
This will be used later in the example of custom noise.

\begin{figure}[h]
    \centering
    \includegraphics[width=0.32\textwidth]{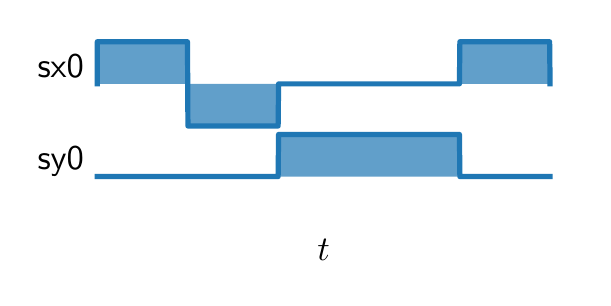}
    \caption{The compiled pulse of a $\pi/2$ pulse followed by a Z gate for the customized processor defined in \cref{sec:customization appendix}. The Z gate is decomposed into rotations over the $x$ and $y$ axes.}
    \label{fig:customize pulse}
\end{figure}

We then initialize a \texttt{ModelProcessor} with this model.
In the \texttt{ModelProcessor}, the default simulation workflow is already defined, such as the \texttt{load\_circuit} method.
Since rotations around the $x$ and $y$ axes are the native gates of our hardware, we define them in the attribute \texttt{native\_gates}.
Providing this native gates set, rotation around $z$ axis will be automatically decomposed into rotations around $x$ and $y$ axes.
We define a circuit consisting of $\pi/2$ rotation followed by a Z gate.
The compiled pulses are shown in \cref{fig:customize pulse}, where the Z gate is decomposed into rotations around $x$ and $y$ axes.

Next, we show an example of defining customized noise and simulating classical cross-talk:
\inputpython{code/code3_customize.py}{143}{219}

In the code block above, we first define a custom \texttt{ClassicalCrossTalk} noise object that uses the \texttt{Noise} class as the base.
The \texttt{get\_noisy\_dynamics} method will be called during the simulation to generate the noisy Hamiltonian model.
Here, we define a noise model that adds the same driving Hamiltonian to its neighbouring qubits, with a strength proportional to the control pulses strength applied on it.
The detuning of the qubit transition frequency is simulated by adding a $\sigma_z$ drift Hamiltonian to the processor, with a frequency of $1.852$ MHz.

Second, we define a random circuit consisting of a sequence of $\pi$ rotation pulses with random phases.
The driving pulse is a $\pi$ pulse with a duration of $25 \, \mu\rm{s}$ and Rabi frequency $20$ KHz.
As described in \cite{piltz2014trapped}, this randomized benchmarking protocol allows one to study the classical cross-talk induced decoherence on the neighbouring qubits.
The two qubits are initialized in the $\ket{00}$ state with a fidelity of 0.975.
After the circuit, we measure the population of the second qubit.
If there is no cross-talk, it will remain perfectly in the ground state.
However, cross-talk induces a diffusive behaviour of the second qubit and the fidelity decreases.
This simulation is repeated 1600 times to obtain the average fidelity, as shown in \cref{fig:cross-talk} in the main text.

\bigbreak

\twocolumn
%

\end{document}